\pdfoutput=1
\documentclass[a4paper]{JACoW-GSI-2013}
\usepackage[utf8]{inputenc}
\usepackage[USenglish]{babel}
\usepackage{graphicx}
\usepackage{textcomp,siunitx} 

\begin{document}

\title{In Situ Measurement of Mechanical Vibrations\\ of a 4-Rod RFQ at GSI}

\author{P. Gerhard, L. Groening, K.--O. Voss\\
GSI Helmholtzzentrum für Schwerionenforschung GmbH, Darmstadt D-64291, Germany\thanks{p.gerhard@gsi.de}}

\maketitle

\begin{abstract}
A new 4-rod RFQ was commissioned at the UNILAC in 2009, it went into operation in 2010 \cite{GER10,GER12}. At high rf amplitudes strong modulations of the rf reflection emerge. They are attributed to mechanical oscillations of the rods, excited by the rf pulse. As these modulations could so far only be seen \emph{during} the rf pulse by means of rf measurements, a direct observation of the mechanical vibrations was desirable. Such measurements have been conducted using a commercial laser vibrometer, allowing the investigation of the mechanical behavior of the RFQ \emph{in situ} and independent of the presence of rf power. Results from investigations under standard operation conditions confirm the vibrations as the source of the modulations observed by rf as well as their excitation by the rf pulse. Further measurements revealed more details about the excitation process, leading to a better understanding and possibly new ways of mitigation.
\end{abstract}

\section{Introduction}
A new 4-rod Radio Frequency Quadrupole RFQ was commissioned at the High Charge State Injector HLI at the UNILAC in 2009. It is in operation since 2010 \cite{GER10,GER12}. At high rf amplitudes strong modulations of the rf reflection emerge, with a modulation frequency of approximately 500 Hz. Such modulations of the reflected rf power were already observed long ago at the first HLI RFQ \cite{VINZENZ97}, but with the new RFQ the impact on the operation became unacceptable. The high fraction of reflected rf power severely limits the pulse length and amplitude achievable. The frequency control is challenging due to the amplitude of the modulation and its variation with rf power level. Stable operation is hard to achieve. A number of investigations based on rf measurements were conducted. They all suffered from the fact, that the measurements were restricted to the timespan of the rf pulse. A different, rf independent approach was required. The idea of using a laser, which could be directed into the cavity through a vacuum window, to monitor the movement of the RFQ electrodes, lead to the application of a laser vibrometer, which allows for remote, fast measurement of the movement of any (reflective) surface. 

\section{Principle of measurement}
The \textsc{Polytec} OFV-525 laser vibrometer exploits the Doppler effect of a laser beam reflected by the surface of interest. The laser beam is split into a probe and a reference beam. The Doppler frequency shift of the reflected light with respect to the reference beam, which is proportional to the velocity of the movement, is determined by a laser interferometer. In order to define the direction of movement relative to the vibrometer, the reference beam light frequency is modulated. The displacement of the surface, and hence the amplitude of the vibration, can either be measured directly by a second operation mode of the vibrometer, or calculated from the velocity data. In principle, only oscillations in the direction of the laser beam can be monitored. For more details about the principles of vibrometry refer to \cite{PolytecWeb}. 

By pointing the laser through a vacuum window onto e.\ g.\ the rounded back edge of a rod between two mounting bridges, we were able to measure the vibration of the rods \emph{in situ} for different operational states of the rf (on; off; different pulse lengths, amplitudes and repetition rates; during the rf pulse or in between two pulses etc.). The laser beam was pointing at the electrode at about 30° (60°) with respect to the principle axes of the electrode. This allowed for the measurement of oscillation modes along both axes. The vibrometer was placed on top the RFQ, and the laser beam was guided into the cavity by a motorized mirror. The beam spot was monitored by a video camera. By this, also other parts of the structure like the mounting bridge or coupling loop could be examined. Reference data were taken in order to exclude any perturbation of the vibrometer by the rf, X-rays, vibration of the mounting or the vacuum window.

\section{Results}
\begin{figure}[htb]
   \centering
   \includegraphics[width=82.5mm] {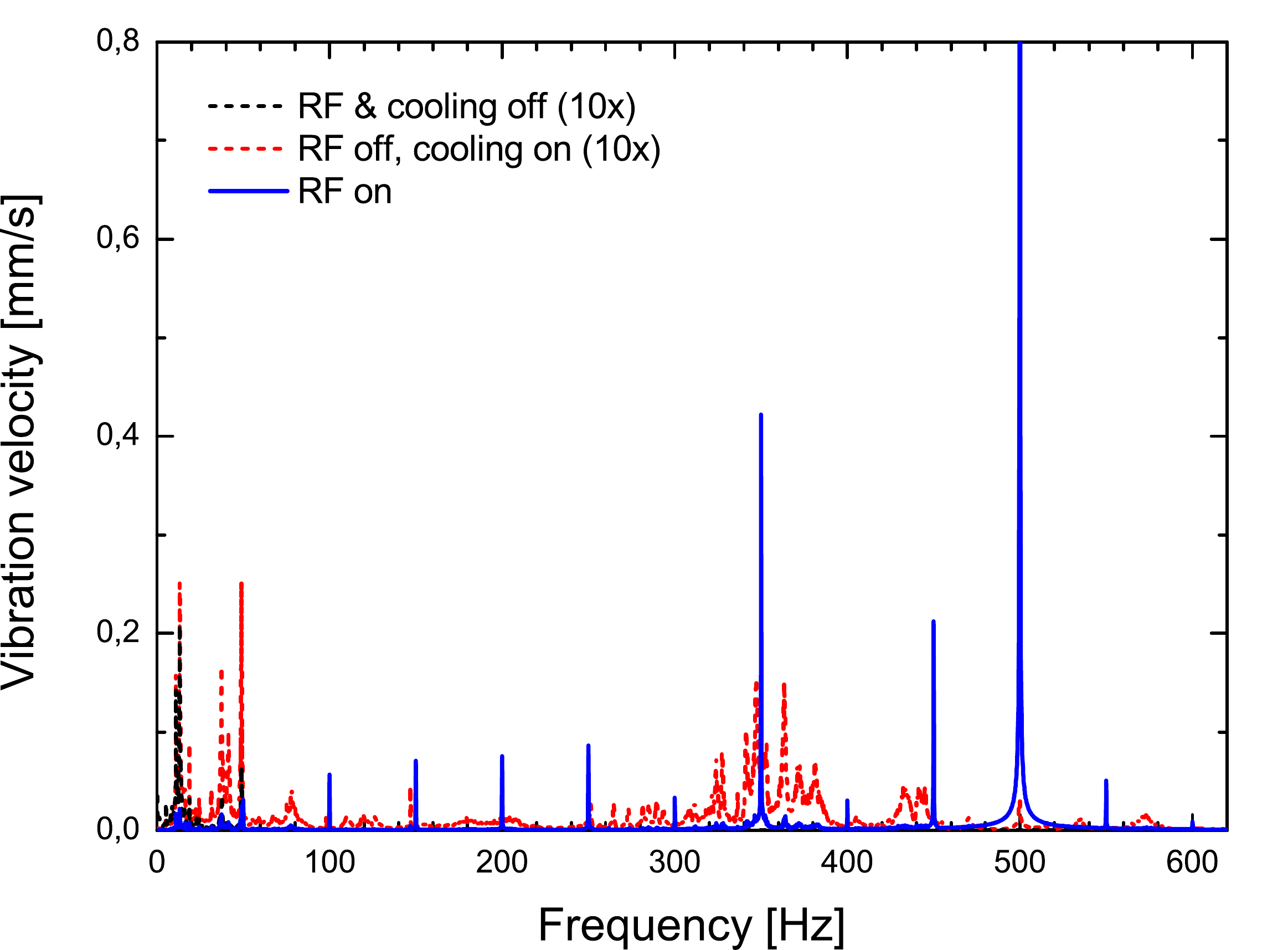}
   \caption{Frequency spectra of the measured vibration velocity. Black: Vacuum pumps running, cooling water and rf switched off. Red: With cooling water. Blue: With RF on at 50~Hz pulse repetition rate. (Black and red are 10x enhanced.)}
   \label{figRFoffon}
\end{figure}

In Fig.\ \ref{figRFoffon}, measurements without rf, but with vacuum pumps and cooling water running, are shown as frequency spectra of the vibration velocity (black and red dashed curves, ten times enhanced). 
The vacuum pumps and other devices lead to vibrations of the electrode rods in the range from a few to about 100~Hz, with noticeable signals around 12 and at 50~Hz. Switching on the cooling water, which runs directly through the electrodes and other parts of the structure, adds a broad background up to 600~Hz, with higher amplitudes between 300 and 400~Hz and below 50~Hz.

The third curve (blue solid) shows a frequency spectrum of the vibration velocity with rf power at moderate levels. The data were taken unsynchronized at a pulse repetition frequency of 50~Hz, and the spectrum is shown as calculated by the vibrometer software. This leads to artefacts in the FFT (narrow side bands every 50~Hz). Besides this, two strong peaks at 500 and 350~Hz emerge when rf power is transmitted. The 500~Hz peak is in very good agreement with the rf observations made earlier and clearly identifies the rf pulses as the source of the vibrations which modulate the rf signals. The peak at 350~Hz was not known from the rf measurements, details will be discussed later. The velocity of the vibration at 500~Hz corresponds to an oscillation amplitude on the order of 1~$\mu$m.

\subsection{Results from Standard Operation Conditions}
\begin{figure}[tb]
   \centering
   \includegraphics[angle=90,width=82.5mm]{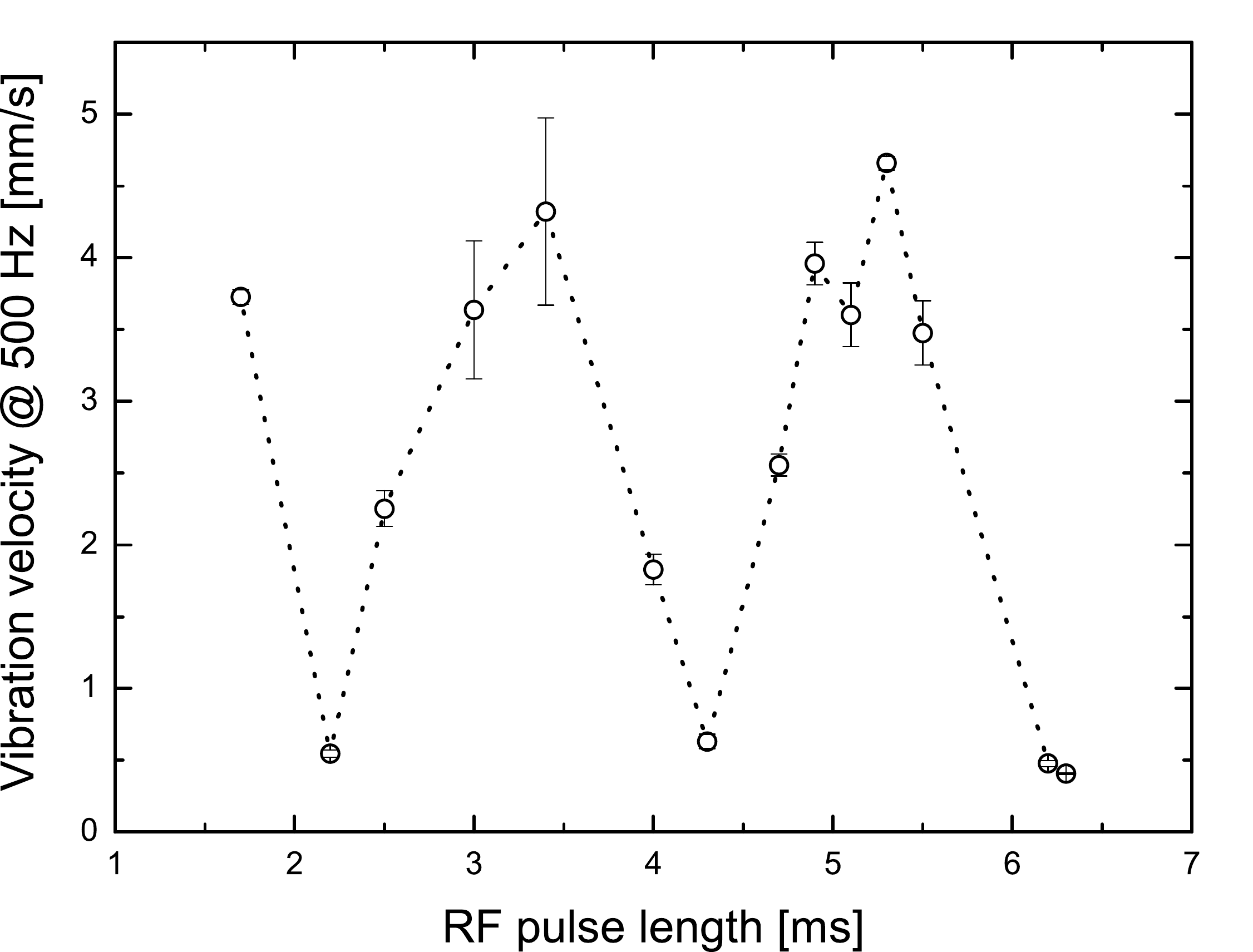} 
   \caption{Vibration velocity at 500 Hz for different rf pulse lengths at 50 Hz pulse repetition rate.}
   \label{figPulseLength}
\end{figure}

After this first affirmative result, specific investigations on the correlation between the excitation and operational parameters, like pulse length and repetition rate, were conducted. Figure \ref{figPulseLength} shows the variation of the vibration velocity at 500~Hz for different rf pulse lengths at 50~Hz pulse repetition rate. 
One can immediately see the periodicity of the velocity as a function of the pulse length. From operational experience and rf observation it was known, that at rf pulse lengths of roughly 2.3, 4.3 and 6.3~ms, corresponding to 1.15, 3.15 and 5.15~ms beam pulse length respectively, stable operation was possible, while at other pulse lengths, only unstable or no operation at all was possible, depending on the power level. This observation corresponds nicely to the periodicity and the position of the minima of the vibration velocity in Fig.\ \ref{figPulseLength}.

\begin{figure}[tb]
   \centering
   \includegraphics[angle=0,width=82.5mm]{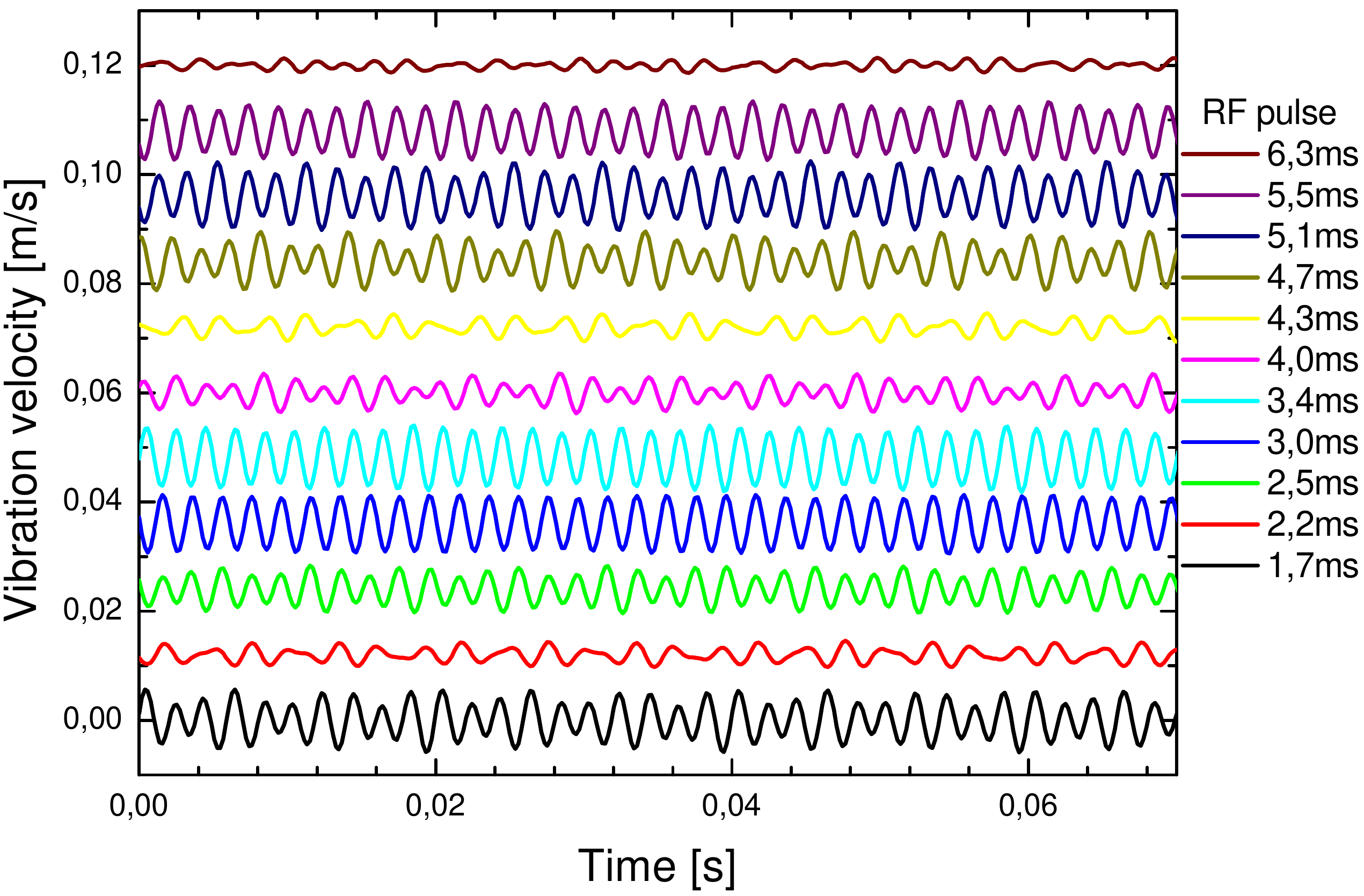} 
   \caption{Comparison of the vibration velocity as a function of time for different rf pulse lengths at 50 Hz repetition rate.}
   \label{figTimeSignals}
\end{figure}

Figure \ref{figTimeSignals} shows the same data as Fig.\ \ref{figPulseLength}, but in the time domain.
The diagram spans 3.5 rf cycles. For any pulse length, a strictly periodic progression of the velocity arises, with a period of at most 20~ms, corresponding to the repetition rate of 50~Hz. The data show, that for the pulse lengths where stable operation is possible, the vibration amplitude is low not only during the rf pulse, but all the time.

\subsection{Further Measurements}
\begin{figure}[tb]
   \centering
   \includegraphics[angle=0,width=82.5mm]{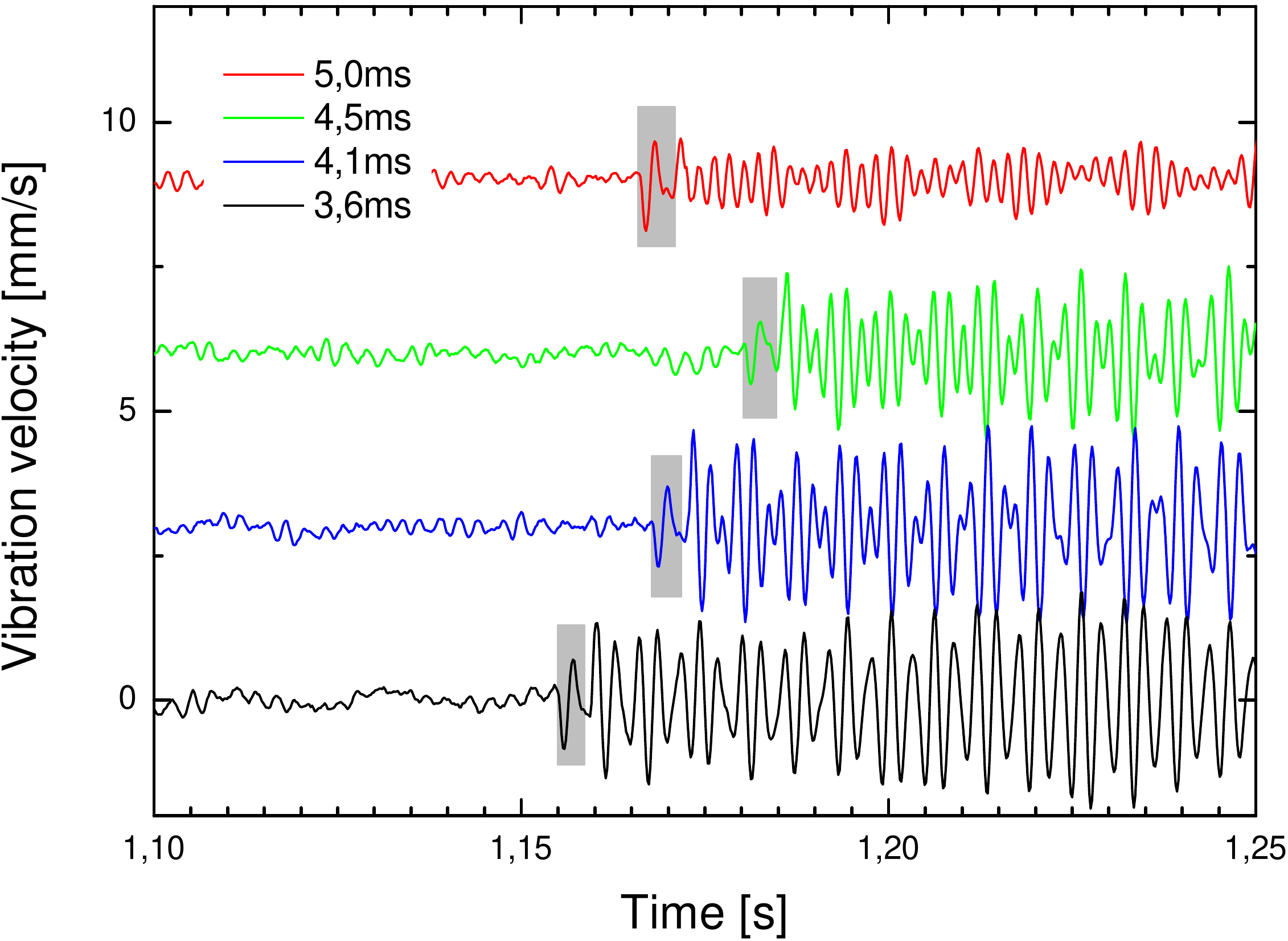} 
   \caption{Vibration velocity as a function of time for different rf pulse lengths, measured at 1~Hz pulse repetition rate. The grey fields mark the time of rf transmission.}
   \label{figTimePuls1Hz}
\end{figure}

\begin{figure}[tb]
   \centering
   \includegraphics[angle=0,width=82.5mm]{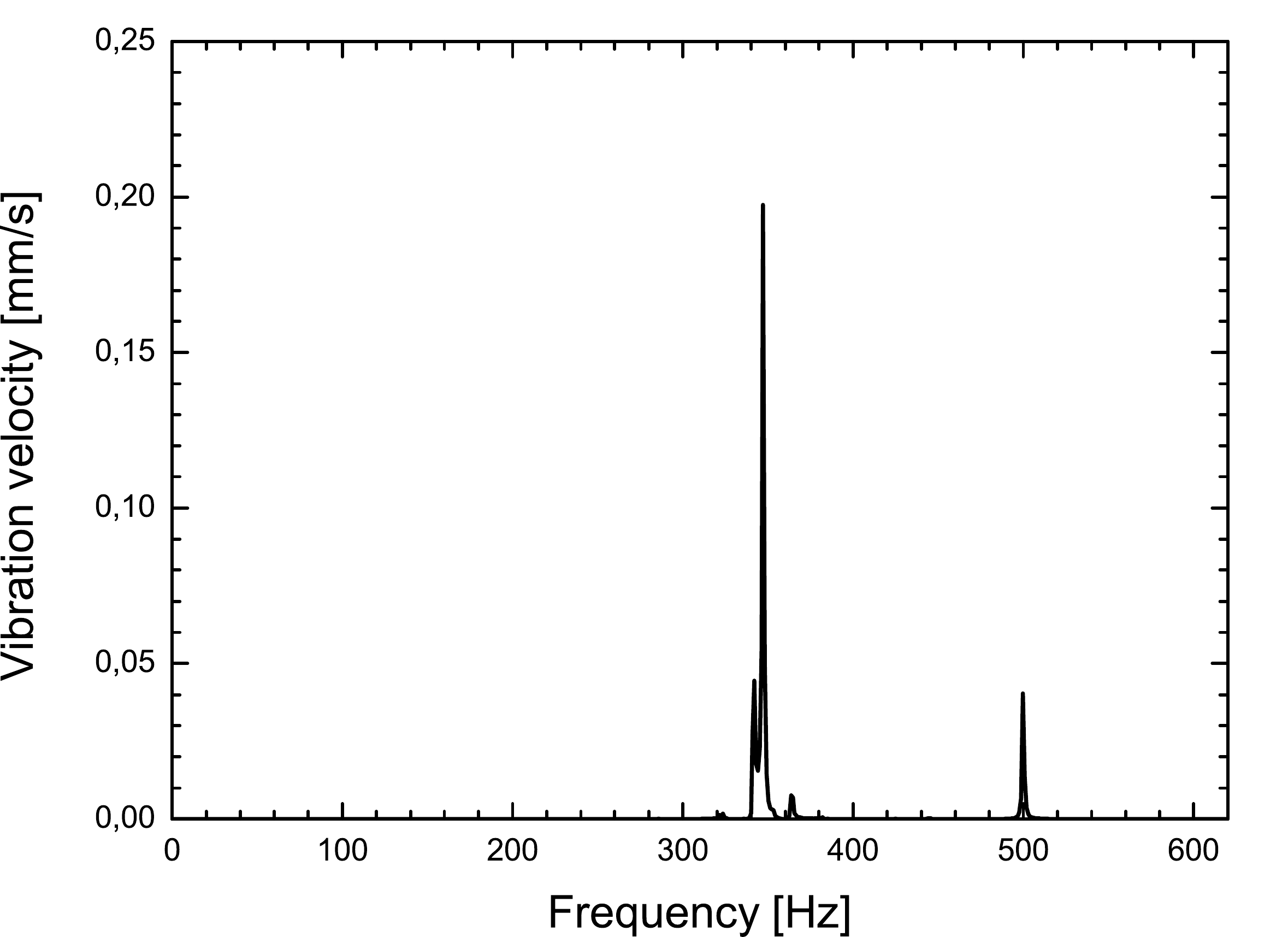} 
   \caption{Frequency spectrum of the measured vibration velocity at 1~Hz rf pulse repetition rate. Only the free decay of the vibration after the rf pulse is considered.}
   \label{figRingdown}
\end{figure}

The vibration in between two rf pulses is not accessible by rf observations. As a substantial improvement, the laser vibrometer enables the continuous analysis of the vibration. This is of high interest, because it allows to study the interference of the vibrations excitated by the two edges of the rf pulse. 
In Fig.\ \ref{figTimePuls1Hz} and \ref{figRingdown} data of the vibration during rf transmission and the free decay, taken at 1~Hz pulse repetition rate, is shown.

Due to the long break between two pulses, the vibration has nearly vanished before a new rf pulse is transmitted, as indicated by the grey field (Fig.\ \ref{figTimePuls1Hz}). The start of the vibration, showing always the same phase and amplitude for the first period, clearly indicates when the rf pulse excites the vibration. However, the exact time where the rising edge of the rf pulse is to be placed can not be determined without synchronization of the vibrometer to the rf cycle, which was not available at the time of these measurements. 

At a time period after the beginning of the vibration, which corresponds approximately to the length of the rf pulse, a phase jump occurs, which most probably marks the second excitation by the falling edge of the rf pulse. As this excitation interferes with the existing vibration from the first edge, the vibration changes its amplitude to higher values at 3.6 and 4.1~ms, while at 5.0~ms the amplitude is considerably lower after the pulse than within.

The laser vibrometer also makes the ring--down of the free mechanical oscillation --- after the excitation by both edges of the rf pulse --- open to investigations. Using a slow repetition rate, it also allows for observations for a longer period of time, and by this an exact analysis of the spectrum and damping times is possible. The frequency spectrum in Fig.\ \ref{figRingdown} shows four major modes at 341, 347, 364 and 500~Hz. The decay times for all frequencies are about 0.3~s. The vibration at 347~Hz dominates the spectrum, while the mode at 500~Hz is the one, that affects the rf. Nevertheless, the other vibration modes may well affect the beam and are therefore also undesirable.

\section{Discussion}
The initial objective of the measurements described above was to verify independently, that 
\begin{Itemize}
\item the rf pulses are the source of the excitation of the vibrations, and
\item these mechanical oscillations are responsible for the modulation of the rf properties.
\end{Itemize}
These results were achieved quickly and without doubt. 

The enhanced opportunities of the laser measurement technique enabled further investigations of the properties of the vibrations. One important result is the determination of the decay times. From the rf observations, decay times on the order of 10~ms were derived, owed to the fact, that only data from the short time span of the rf pulse was available \cite{GER12}, which are additionally distorted by the rising edge of the rf pulse. The new data, taken at 1~Hz pulse repetition rate, give decay times on the order of 0.3~s. Together with the analysis of the time resolved data of the movement at 50~Hz pulse repetition rate, the explanation of the different behavior at different pulse repetition rates given in \cite{GER12} has to be revised. From Fig.\ \ref{figPulseLength} it shows, that the perturbation within the rf pulse is minimal for pulse lengths $t_{P}$ around 2~ms and integer multiples of this, which is exactly the period $t_V$ of the mechanical vibration. The excitation of the rods by the electrical force due the rf voltage changes sign for the rising and falling edge of the pulse. When $t_{P}=n\cdot t_V,\; n=1,2,\ldots\;$, the two excitations match the same phase of the vibration and therefore cancel out. When $t_{P}=\frac{2n+1}{2}\cdot t_V, n=0,1,\ldots$, both the excitations and the vibration phase change sign and therefore interfere positively in this case. Figure \ref{figTimeSignals} reveals, that in the first case, the oscillation amplitude is always low and not only during the pulse, as assumed before. There is no particular strong mechanical oscillation present at the beginning of the rf pulse, which could cancel out the excitation from the rising edge. It seems to be merely a question of phase. Synchronization between the rf cycle and the vibrometer measurements is necessary to explore this in more detail.

The direct measurement of the mechanical movement of the electrodes also revealed, that different oscillation modes are excited. Some of which have an effect on the electrical properties of the structure, while others have not. This can be explained by different orientations of the oscillation modes with respect to the principal axes of the structure. The latter modes are not discoverable by rf measurements, but are still to be considered because they may well affect the beam dynamical properties of the accelerating structure. 

\section{Summary \& Outlook}
The use of a laser vibrometer allowed a detailed investigation of the mechanical properties of a 4-rod RFQ. Results from investigations under standard operation conditions confirmed the vibrations as the source of the modulations observed by rf as well as their excitation by the rf pulse. Especially the observation of the movement independent of the rf pulse revealed the interference scheme. This lead to a better understanding of what was known from operational experience before.

In order to understand the connection between the vibrations and their impact on the rf and other properties in more detail, simulations of the vibrational modes and their effect on the electrical parameters of the structure are initiated and have to be extended. The main goal will be to identify the vibrational mode at 500~Hz, which perturbates the rf properties and impairs stable operation of the RFQ. Based on this, strategies to mitigate the problems have to be developed, e.\ g.\ a new set of electrodes with a modified profile, which is less susceptible to the vibrations.

The knowledge gained by the measurements described here may also open easier ways to reduce the vibrations. One way could be to modify the existing electrodes only in some parts, for instance the long free ends. A completely different approach is to modify the rf pulses in such a way, that the vibrations within the accelerating part of the pulse are cancelled out. This requires a better understanding of how the rf pulses and the electrodes interact, which can be provided only by the vibrometer data. Further measurements with synchronization to the rf cycle could help analysing the details of the excitation process. Such investigations should be accompanied by implementing new rf pulsing schemes.


\end{document}